\documentstyle[emulateapj,psfig]{article}

\newcommand{\be}{\begin{equation}}
\newcommand{\ee}{\end{equation}}
\newcommand{\ba}{\begin{eqnarray}}
\newcommand{\ea}{\end{eqnarray}}
\newcommand{\siml}{\lower4pt \hbox{$\buildrel < \over \sim$}}
\newcommand{\simg}{\lower4pt \hbox{$\buildrel > \over \sim$}}

\slugcomment{2003, ApJ, V595, in press}

\begin{document}

\title{High Energy Neutrinos from Magnetars}

\author{Bing Zhang$^1$, Z. G. Dai$^{2,1}$, P. M\'esz\'aros$^{1,3}$
E. Waxman$^{4}$ and A. K. Harding$^{5}$}

\affil{$^1$Dept. of Astronomy \& Astrophysics, Pennsylvania
State University, University Park, PA 16802 \\
$^2$Dept. of Astronomy, Nanjing University, Nanjing 210093,
P. R. China \\
$^3$Dept. of Physics, Pennsylvania State University, University Park,
PA 16802 \\
$^4$Dept. of Condensed Matter Physics, Weizmann Institute
of Science, Rehovot 76100, Israel \\
$^5$Laboratory of High Energy Astrophysics, NASA Goddard Space Flight
Center, Greenbelt, MD 20771}


\begin{abstract}
Magnetars can accelerate cosmic rays to high energies through the
unipolar effect, and are also copious soft photon emitters. We show
that young, fast-rotating magnetars whose spin and magnetic moment
point in opposite directions emit high energy neutrinos from their
polar caps through photomeson interactions. We identify a neutrino
cut-off band in the magnetar period-magnetic field strength phase
diagram, corresponding to the photomeson interaction threshold. Within 
uncertainties, we point out four possible neutrino emission candidates
among the currently known magnetars, the brightest of which may be
detectable for a chance on-beam alignment. Young magnetars in the 
universe would also contribute to a weak diffuse neutrino background, 
whose detectability is marginal, depending on the typical neutrino 
energy. 
\end{abstract}



\section{Introduction}

The most widely discussed high energy neutrino sources include
gamma-ray bursts (e.g. Waxman \& Bahcall 1997, 2000; Dai \& Lu 2001;
M\'esz\'aros \& Waxman 2001; Guetta, Spada \& Waxman 2001), blazars
(e.g. Stecker et al. 1991; Atoyan \& Dermer 2001) and micro-quasars
(e.g. Levinson \& Waxman 2001). Another type of objects, pulsars, 
have been also considered for some time to be high energy neutrino
emitters (e.g. Eichler 1978; Helfand 1979).
The direct motivation is that pulsars are unipolar generators which
induce a large potential to accelerate protons to high energies
(see Blasi, Epstein \& Olinto 2000; Arons 2003, for recent 
discussions on the topic). Such high energy protons,
when colliding with target photons or materials, can generate
neutrinos mainly through pion decay. In the immediate environment
of a pulsar (e.g. within its magnetosphere), however, the
main difficulty is the lack of a large enough target column 
density for pion production. 
The pulsar neutrino emission is therefore usually discussed within 
the context of pulsar wind nebulae (Beall \& Bednarek 2002; 
Bednarek 2001) or binary systems (e.g. Eichler 1978). 

Here we show that the conditions for neutrino production via 
photomeson interaction are realized in the inner magnetospheres 
of the so-called ``magnetars'', pulsars with superstrong surface 
magnetic fields ($B_s \sim 10^{15}$ G) (Duncan \& Thompson 1992; 
Pacz\'ynski 1992; Usov 1992), if they are young enough to allow 
acceleration of protons above the photomeson threshold. These objects 
are the leading candidate for explaining the widely observed soft 
gamma-ray repeaters (SGRs) and anomalous X-ray pulsars (AXPs) 
(Thompson \& Duncan 1995, 1996; see observational reviews by Hurley
2000, Mereghetti 2000, and a theoretical review by Thompson 2000).
We will estimate the high energy neutrino flux of these objects, as
well as their detectability with currently operating and planned large
area neutrino telescopes, such as AMANDA-II, ICECUBE, ANTARES, NESTOR
and NEMO. We will also estimate a diffuse high energy neutrino
background contributed by all young magnetars in the universe.

\section{Basic picture and photomeson threshold}

In the original definition of Thompson \& Duncan, magnetars are
neutron stars powered by decaying magnetic fields (Goldreich \&
Reisenegger 1992; Thompson \& Duncan 1996; Heyl \& Kulkarni 1998;
Colpi, Geppert \& Page 2000). However, since they are rotating neutron
stars, a conventional power source, i.e., the spindown energy of the
star should also play a noticeable role. Although in the slowly
rotating magnetars (corresponding to SGRs and AXPs) the spindown power
is much lower than the magnetic power, it can exceed the magnetic
power during the 
magnetar's early life time when the star spins much more rapidly.
In any case, there are in principle two main energy sources that 
power a magnetar. Within the context of our proposed neutrino
production mechanism, the dominant photomeson interaction leading to 
neutrinos occurs through the $\Delta$-resonance,
\be
p\gamma \to \Delta \to n \pi^{+} \to n \nu_\mu \mu^{+} \to n
\nu_\mu e^{+} \nu_e \bar\nu_\mu. 
\label{pgam}
\ee
In magnetars, the spindown power serves to accelerate protons, and
the magnetic power provides copious near-surface photon targets,
so that the condition for photomeson interaction is in principle
realized. 
In order to achieve substantial neutrino production rate, however,
a threshold condition 
\begin{equation}
\epsilon_p \epsilon_\gamma \simg 0.3 ~({\rm
GeV})^2 f_g
\label{th}
\end{equation}
has to be satisfied. Here 
\be
f_g \equiv (1-\cos\theta_{p\gamma})^{-1}
\label{fg}
\ee 
is a geometric factor, where $\theta_{p\gamma}$ is the maximum
lab-frame incidence angle between protons and photons.

The maximum potential drop of a magnetar with the rotation
frequency, $\Omega=2 \pi/P$ (where $P$ is the period), and surface
magnetic field at the pole, $B_p=10^{15}{\rm G} B_{p,15}$, is
\begin{equation}
\Phi=\frac{\Omega^2 B_p R^3}{2 c^2} \simeq 6.6\times 10^{15}~{\rm V}~
B_{p,15} R_{6}^3 P^{-2}~,
\label{Phi}
\end{equation}
where $R=10^6{\rm cm}R_6$ is the stellar radius.
For a fraction $f=0.5 f_{1/2}$ (the nominal value is for the 
case assuming a random distribution of the inclination angles) 
of the magnetars whose spin and magnetic moment point in opposite 
directions, i.e., ${\bf \Omega} \cdot {\bf B}_p < 0$, the 
spin-induced parallel electric field accelerates positive charges 
from the surface.
The neutron star surface composition is poorly known. Current
modeling of the X-ray data from some nearby isolated neutron stars
suggests that it likely consists of light elements such as hydrogen or
helium (Zavlin, Pavlov \& Tr\"umper 1998; Sanwal et al. 2002) rather
than heavy elements such as iron. Recently, Ibrahim et al. (2002,
2003) reported discoveries of the cyclotron resonance features from 
SGR 1806-20 during outburst which are well interpreted as proton
cyclotron features in a magnetar environment, lending credence that
the surface composition of magnetars is hydrogen. Here we assume 
a hydrogen composition of the magnetar surface. According to the
standard pulsar theory, protons are accelerated from the polar cap
region within charge-depleted gaps (e.g. Ruderman \& Sutherland 1975;
Arons \& Scharlemann 1979). If surface temperatures of magnetars are
high enough to allow free emission of protons, space-charge limited
flow models of electron acceleration (Arons \& Sharlemann 1979;
Harding \& Muslimov 1998, 2001, 2002) will apply to the acceleration
of protons in this case.

Magnetars are also strong X-ray emitters. During the early epochs of
their lives, magnetars emit thermal radiation, thought to be due to
decay of the strong magnetic fields. This maintains 
a high X-ray luminosity (typically $10^{35}-10^{36} {\rm erg~s^{-1}}$ 
as observed in SGRs/AXPs) over a long period of time. 
The nominal field decay law, $d B_p/d t=-b B_p^{1+\delta}$, leads  
to a magnetic field strength time dependence $B_p(t) = {B_{p,0}}/ 
{[1+t/\tau_{mag}]^{1/\delta}}$ (Colpi et al. 2000), 
where $B_{p,0}$ is the initial surface field at the pole. Thus the 
magnetar surface field, and hence its quiescent X-ray luminosity, 
remains almost constant for a typical decaying timescale
$\tau_{mag} = (b \delta B_{p,0}^\delta)^{-1} \sim 10^4$ yr.
The observed blackbody temperature for SGR/AXP quiescent emission
is $k T_\infty \sim (0.4-0.6)$ keV (e.g. Hurley 2000; Meregheti 2001;
Thompson 2001). The typical near-surface photon energy is therefore
\be
\epsilon_\gamma = 2.8 k T_\infty (1+z_g) \sim (1.6-2.4)~{\rm keV},
\label{epsgam} 
\ee
where $(1+z_g) \sim 1.4$ is the near-surface gravitational redshift.
The threshold proton energy from eq.(\ref{th}) is therefore
\begin{equation}
\epsilon_{p,th} \sim (125-188) f_g~{\rm TeV}~.
\label{eps_th}
\end{equation}

The potential across the polar cap [eq.(\ref{Phi})] drops as $P^{-2}$
as the star spins down. The typical energy of the protons accelerated in
the inner gap is a fraction of $e\Phi$, which also drops as the magnetar
ages. The proton energy acquired in the inner gap can be written
as $\epsilon_{p} = \eta_p e\Phi$, where $\eta_p$ parameterizes the
uncertainties in the utilization of the polar cap unipolar potential.
The maximum efficiency could be as high as $\eta_{p,max} \sim
(0.15-0.85)$ (eqs.[2] and [7] of Zhang, Harding \& Muslimov 2000),
if the electric field parallel to the magnetic field line is not
effectively screened, and if the particle acceleration is not
radiation-reaction-limited.
For proton accelerators, radiation reaction is negligible for the 
known magnetar candidates (SGRs/AXPs), mainly because of the small
magnetic field curvature in slow rotators. Soft X-rays could be 
Lorentz-boosted in the proton's rest frame and pair-produce in its 
Coulomb field (Cheng \& Ruderman 1977; Heitler 1954),
with a cross section $\sigma_{\pm} \sim Z^2 \alpha \sigma_{T}
(3/8 \pi) [(28/9) \ln (E'_\gamma/m_e c^2)-218/27] \sim Z^2 \cdot
7\times 10^{-27}~{\rm cm}^2$ for $\epsilon_p \sim \epsilon_{p,th}$,
where $Z$ is atomic number, $\alpha$ is the fine-structure constant,
$\sigma_{T}$ is the Thomson cross section, and $E'_\gamma$ is the
Lorentz-boosted photon energy in the rest frame of the proton. 
The typical pair-production mean
free path is $l_{\pm} \sim (n_\gamma \sigma_{\pm})^{-1} \sim
1.3\times 10^4$ cm, comparable to that in old normal pulsars whose 
primary pairs are produced through non-resonant inverse Compton 
scattering. Comparing with the numerical results in the pulsar 
case (Harding \& Muslimov 2002), the pairs produced in such an
environment are likely to be too few to fully screen the parallel 
electric field before the protons reach the photomeson threshold.
The cascade pair multiplicity for old magnetars is small. 
As a result, $\eta_p \sim \eta_{p,max}$ could be achieved in
old magnetars, so that
\begin{equation}
\epsilon_{p,max} \simeq \eta_{p,max} e\Phi
\simeq (40-220)~{\rm TeV}~ B_{p,15} R_6^3 (P/5~{\rm s})^{-2}~.
\label{eps_pmax}
\end{equation}
Although in young magnetars the parallel electric fields may be
screened before $\epsilon_p$ reaching $\epsilon_{p,max}$, this
nonetheless happens well above $\epsilon_{p,th}$. Thus, one can
define a ``neutrino death valley'' for magnetars by
requiring that $\epsilon_{p,max}$ (\ref{eps_pmax}) exceeds the
threshold energy $\epsilon_{p,th}$ (\ref{eps_th}), which gives
\begin{equation}
P < (2.4-6.8)~{\rm s}~ B_{p,15}^{1/2} R_6^{3/2} f_g^{-1/2}~.
\label{valley}
\end{equation}
The range of periods in the right hand side of eq.(\ref{valley})
defines two diagonal lines in the $P-B_p$ plane (Figure 1).
Photomeson interactions and neutrino emission cease when the
magnetar crosses this valley from left to right during its evolution.
Magnetars lying in the valley itself are marginal neutrino emitters,
i.e., they could be neutrino-loud for favorable parameters.

\section{Discrete sources}

Here we investigate the possibility of detecting neutrino emission 
from the individual known magnetar candidates (i.e. SGRs and
AXPs), which are typically slow rotators. 

As seen in eq.(\ref{valley}), the location of the neutrino death
valley depends on the geometric factor $f_g$ (eq.[\ref{fg}]). Below we
will discuss the possible value of this factor. For the simplest case, 
thermal photons are expected to be emitted from the surface
semi-isotropically, so that $\theta_{p\gamma} \leq 90^{\rm o}$ and
$f_g \geq 1$. 
In recent magnetar models (Thompson, Lyutikov \& Kulkarni 2002)
the magnetosphere is assumed to be globally twisted and 
current-carrying. The non-relativistic charges in the closed field
line region form a resonant cyclotron screen at a high altitude (about 
10 stellar radii) with an optical depth higher than unity (see also
Wang et al. 1998 for a similar discussion within the context of normal
pulsars). The emergent X-ray photons 
would endure multiple Comptonization before escaping, and the mechanism
is used to interpret the observed hard X-ray non-thermal tail in the
SGR/AXP spectrum (Thompson et al. 2002). In such a picture, it is
natural to expect some downward X-ray photons (with a luminosity
comparable to what is observed) reflected from the resonant 
cyclotron screen into the open field line region. Here we assume a
$\sim 50\%$ 
efficiency of backscattering of the surface thermal photons, but the
real fraction has to be treated more carefully by incorporating the
detailed radiation transfer processes. In such a most favorable case, 
$\theta_{p\gamma} \siml 180^{\rm o}$ could be achieved so that $f_g
\simg 1/2$. This is the most optimistic case for neutrino production.

If such a pair reflection screen is ineffective, however, $f_g$ is
larger. Defining the typical photomeson interaction mean free
path as $l_{p\gamma}$ and the physical hot spot radius as $r_h$,
one can estimate\footnote{The length for the proton to achieve
significant acceleration is at most comparable to $l_{p\gamma}$ within
various acceleration models (Harding \& Muslimov 1998; Ruderman \&
Sutherland 1975). Therefore we may regard $l_{p\gamma}$ as the typical 
vertical length scale for photomeson interaction.} 
\begin{equation}
\cos \theta_{p\gamma} = \left\{
\begin{array}{@{\,}ll}
\left[1-\left(\frac{R}{R+ l_{p\gamma}}\right)^2\right]^{1/2}, &
l_{p\gamma} \leq l_{cr}, \\  
\left[1+\left(\frac{r_h}{l_{p\gamma}}\right)^2\right]^{-1/2}, &
l_{p\gamma} > l_{cr}.
\end{array}
\right.
\label{cos}
\end{equation}
where $l_{cr}$ satisfies the relation $[1+(r_h/l_{cr})^2]
[1-(R/R+l_{cr})^2=1$, and is the critical height at which the horizon
is just the boundary of the hot spot. In typical SGRs/AXPs, the
surface area of the hot region is large, which could be roughly
estimated as $A\sim L / \sigma [(1+z_g)T_\infty]^4\sim 4.0\times
10^{11}~{\rm cm}^2~ L_{35} (kT_\infty / 0.5{\rm keV})^{-4}$, where
$\sigma=ac/4$ is the Stefan-Boltzmann constant. This gives a rough
estimate of the ``hot spot'' radius $r_h \sim 3.6\times 10^5~{\rm cm}~
L_{35}^{1/2} (kT_\infty/0.5{\rm keV})^{-2}$. Given a typical neutron
star radius $R \sim 10^6$ cm and this particular $r_h$ value, one can
solve for $l_{cr} \sim 3.0\times 10^5$ cm. On the other hand, the
photomeson interaction mean free path could be estimated as
$l_{p\gamma}\simeq (n_\gamma  \sigma_{p\gamma})^{-1} \simeq 1.8\times
10^5~{\rm cm}~ (kT_\infty/0.5{\rm keV})^{-3}$, where $n_\gamma \sim
(a/2.8k) [(1+z_g)T_\infty]^3 \simeq 1.1\times 10^{22} ~{\rm cm^{-3}}~
(kT_\infty/0.5{\rm keV})^3$ is the soft photon number density,
$\sigma_{p\gamma} \sim 5\times 10^{-28}{\rm cm}^2$ is the 
photomeson interaction cross section, and $a$ and $k$ are the
blackbody radiation density constant and the Boltzmann constant,
respectively. We can see $l_{p\gamma}< l_{cr}$ for typical parameters,
so that the first expression in eq.(\ref{cos}) is relevant. The
salient feature of this expression is that it does not depend on the
poorly determined parameter $r_h$, as long as it is large enough.
The gravitational bending effect also helps to decrease $f_g$. For the 
above typical parameters, we have $\cos \theta_{p\gamma}=[1-(R/R+
l_{p\gamma})^2]^{1/2} \sim 0.5$, or $f_g \sim 2$. Both $l_{p\gamma}$
and $r_h$ are steep function of the surface temperature, but for
reasonable magnetar surface temperatures, e.g., $kT_\infty > 0.4$ keV,
$l_{p\gamma} < l_{cr}$ is satisfied. However, $f_g$ will be
significantly increased for smaller $kT_\infty$, rendering all
SGRs/AXPs below the neutrino death valley. On the other hand, however, 
higher surface temperatures (which is usually associated with the
post-burst situation) and a larger neutron star radius will decrease
$f_g$ and significantly ease the threshold condition. Observationally
it has been found that the SGR quiescent luminosity is greatly
increased for a long period of time (e.g. Woods 2003), so that we
expect a more facilitated condition and a higher luminosity for a
post-burst magnetar.

In Figure 1, we plot all the SGRs and
AXPs with period $P$ and spin-down rate $\dot P$ measurements relative
to the neutrino death valleys for $f_g=1/2$ (dotted), $f_g=1$ (solid)
and $f_g=2$ (dashed). The polar cap magnetic field for each magnetar
is estimated as $B_p=6.4 \times 10^{19}~{\rm G}~(P \dot
P)^{1/2}R_6^{-3} I_{45}^{1/2}$, where $I$ is the stellar 
moment of inertia. A typical magnetar evolutionary track with
$\tau_{mag} \sim 10^4$ yr is also plotted, with typical ages marked. 
We find that, although none of the magnetars are firmly above any of 
the death valley definitions, four of them are within the 
$f_g \sim 1/2$ valley, two of them are within the $f_g = 1$ valley, 
and SGR 1900+14 lies within the $f_g=2$ valley. 

We estimate the neutrino emission luminosity $L_\nu$ for these
``marginal'' magnetars. If pions decay immediately after production, 
the neutrino emission power can be estimated as $P_\nu \sim 0.05
\cdot (4/3) \sigma_{p\gamma} c \gamma_p^2 aT^4 \sim 4\times 10^5{\rm
erg~s^{-1}} (\epsilon_p / 100{\rm TeV})^2 (T_\infty/0.5{\rm keV})^4$,
where the factor $0.05=(1/4)\eta_{p\to\pi}$, $\eta_{p\to\pi} \simeq
0.2$ is the average fraction of the energy transferred from the proton
to the pion, and the factor (1/4) takes into account the equal
energy distribution among other three leptons besides
$\nu_\mu$ (Halzen \& Hooper 2002). The total number of protons is 
estimated as $N_p \sim n_{gj} \int_0^{r_\nu} A_{o}(r) dr$, where 
$n_{gj}=\Omega B(r)/2\pi ce$ is the Goldreich-Julian (1969) 
number density, $A_o(r)$ is the cross section of the open-field 
line region, and $r_\nu \sim R$ is the typical length for effective 
neutrino radiation. This gives 
\ba
L_{\nu,0} & = & N_p P_\nu \sim 2.3\times 10^{33}~{\rm erg/s} \nonumber 
\\
& \times & \left(\frac{\eta_p}{0.5}\right)^2
B_{p,15}^3 R_6^{10} \left(\frac{P}{5{\rm s}}\right)^{-6}
\left(\frac{T_\infty} {0.5{\rm keV}}\right)^4. 
\ea

The $\pi^{+}$'s can also undergo radiative loss and possible 
reacceleration in the unscreened parallel electric field 
before decaying to $\nu_\mu$. The final neutrino luminosity
is then 
\be
L_\nu=f_c L_{\nu,0}, 
\ee
where $f_c$ is a correction factor 
for cooling or reacceleration. For the four marginal magnetars,
reacceleration is not important, since $\epsilon_p$ is already close
to $\epsilon_{p,max}$. The synchrotron cooling is rapid,
and the $\pi^{+}$'s soon settle into their ground Landau state with a
parallel energy component $\gamma_\parallel=\gamma/(1+(\gamma^2 -1)
\sin^2\theta)^{1/2}$ (Zhang \& Harding 2000), which is essentially
$\sim \gamma$ when the pion injection angle $\theta \sim 0$ (which is
usually valid, especially for slow rotators). For threshold
interactions, the typical pion Lorentz factor upon production is
$\gamma_{\pi^{+}}\sim 2.1\times 10^5 
(\epsilon_{\pi^{+}}/30{\rm TeV})$. 
The pions cool via inverse Compton scattering (IC) with thermal
photons in the Klein-Nishina regime, with a cooling time-scale
$\tau_{\rm IC} \sim 3 \times 10^{-4}~{\rm s}~
(\epsilon_{\pi^{+}}/30{\rm TeV})^{-1} (\epsilon_\gamma/2{\rm keV})^{-4}$,
which is shorter than the pion decay time $\tau_{decay} \simeq
2.6 \times 10^{-8}~{\rm s}~\gamma_{\pi^{+}}=5.5\times 10^{-3}~{\rm s}~
(\epsilon_{\pi^{+}}/30~{\rm TeV})$. Thus the pions undergo some IC cooling
before they decay, and the typical neutrino energy is down by a factor
$f_{c}=(0.3/5.5)^{1/2}\sim 0.23$ relative to the $\epsilon_{\nu,th}$.
Therefore for discrete sources, the typical neutrino energy is
\begin{equation}
\epsilon_\nu \sim 0.05f_{c}\epsilon_{p,th} \sim (1.4-2.2)f_g ~{\rm TeV}.
\label{eps_nu}
\end{equation}
The final neutrino luminosity is 
\ba
L_\nu & = & L_{\nu,0} f_c \sim 5.8\times10^{32} ~{\rm erg/s}
\left(\frac{\eta_p}{0.5}\right)^2 \nonumber \\
& \times & 
\left(\frac{f_c}{0.25}\right)B_{p,15}^3R_6^{10}
\left(\frac{P}{5{\rm s}}\right)^{-6} 
\left(\frac{T_\infty}{0.5{\rm keV}}\right)^4. 
\ea
Since the spindown luminosity is  $L_{sd}=1.5\times 10^{34}
{\rm erg/s}B_{p,15}^2 R_6^6 (P/5{\rm s})^{-4}$, the neutrino
emission efficiency is then
\ba
\eta_\nu & = & L_\nu/L_{sd} \sim 0.04 \left(\frac{\eta_p}{0.5}
\right)^2 \nonumber \\
& \times & \left(\frac{f_c}{0.25}\right) B_{p,15}R_6^4
\left(\frac{P}{5{\rm s}}\right)^{-2}
\left(\frac{T_\infty}{0.5{\rm keV}}\right)^4.  
\ea

We assume that this luminosity is beamed into a sweep-averaged solid
angle $\Delta \Omega_\nu \sim 0.1$, which is typical for a polar cap
angle $\sim 0.01$ and a moderate inclination angle of the rotator.
A smaller/larger $\Delta \Omega_{\nu}$ increases/decreases the on-beam
neutrino flux, but decreases/increases the probability of on-beam
detection. For an on-beam observer, the neutrino number flux at earth
is 
\ba
\phi_\nu &=& \frac{L_\nu}{\Delta \Omega_\nu D^2 \epsilon_\nu}
\sim 2.1\times 10^{-12}~{\rm cm^{-2}~s^{-1}}
\nonumber \\
&\times &\left(\frac{\Delta\Omega_\nu}{0.1}\right)^{-1} 
\left(\frac{\eta_p}{0.5}\right)^2
\left(\frac{f_{c}}{0.25}\right) B_{p,15}^3 R_6^{10} \nonumber \\
&\times &
\left(\frac{P}{{5~{\rm s}}}\right)^{-6}
\left(\frac{T_\infty}{0.5{\rm keV}}\right)^4
\left(\frac{D}{5~{\rm kpc}}\right)^{-2}
\left(\frac{\epsilon_\nu} {2~{\rm TeV}}\right)^{-1},
\ea
where $D$ is the distance to the source. The probability of detecting a
neutrino-induced upward muon with planned neutrino telescopes is
$P_{\nu\to\mu}\simeq 1.3\times 10^{-6} (\epsilon_{\nu}/{\rm TeV})$
(Halzen \& Hooper 2002), giving an on-beam upward muon event rate
\ba
\frac{dN}{dAdt} ({\rm discrete}) \simeq 1.7~{\rm km^{-2}
yr^{-1}} \left(\frac{\Delta\Omega_\nu}{0.1}\right)^{-1}
\left(\frac{\eta_{p}}{0.5}\right)^2\nonumber \\
\times 
\left(\frac{f_{c}}{0.25}\right) B_{p,15}^3 R_6^{10}
\left(\frac{P}{5~{\rm s}}\right)^{-6}
\left(\frac{T_\infty}{0.5{\rm keV}}\right)^4
\left(\frac{D}{5~{\rm kpc}}\right)^{-2}.
\label{dN/dAdt}
\ea
The chances for the observer to be in the neutrino beam are not 
large. Nonetheless, there is a small but finite probability for 
directly detecting some neutrinos from these objects.
In Table 1, we give the predicted muon event rates for the four
magnetar candidates which may be neutrino loud under favorable
conditions, assuming an on-beam observation. Since other magnetars 
all lie below the most favorable $f_g=1/2$ death valley, they are
definetely below the photomeson threshold, and we do not consider 
them as neutrino emission candidates. From Table 1, we see that SGR
1900+14 and 1E 1048-5937 may be detected by km$^3$ 
telescopes with several years of operation, if they are above the 
photomeson threshold and if their neutrino beams sweep the Earth.
Accompanying the neutrinos there should also be electromagnetic 
signals from $\pi^{0}$-decay and $\pi^{+}$ cooling and cascading. 
The high $\gamma B$ and $\gamma\gamma$ pair-formation and 
photon splitting opacity in the strong magnetic fields may degrade 
the typical photon energy to $\lesssim$ 40 MeV, below the EGRET band 
(e.g. Harding, Baring \& Gonthier 1997; Baring \& Harding 2001), but 
it may fall into the INTEGRAL band.

\section{Diffuse flux} 

A direct inference from the above proposal is that the entire 
population of young magnetars in the universe will contribute to
a diffuse neutrino background, before crossing the neutrino death valley. 
The number flux of this background can be generally estimated as
\begin{equation}
\bar\phi_\nu \simeq \frac{0.5 f_{1/2}}{4\pi \bar\epsilon_\nu}
\int_0^{D_H} \left[\int_0^{\tau_{mag,\nu}}
\frac{L_{\nu}(t) f_b(t)}{4\pi f_b(t) D^2} dt \right]
{\cal R}(D) (4\pi D^2)dD~,
\label{barphinu}
\end{equation}
where $D_H \sim 10^{28}$ cm is the Hubble distance, and $\bar
\epsilon_\nu$ is the typical energy of the neutrino background. The
inner integral is the average total neutrino energy fluence per
magnetar emitted towards earth during its neutrino-loud life time
$\tau_{mag,\nu} \sim 5\times 10^3$ yr, which is based on the known
magnetars being marginal neutrino emitters. 
Since 9 magnetars have been discovered in the Galaxy with typical ages
of $10^4$ yr, the local (redshift $z=0$) magnetar birth rate can be 
conservatively estimated 
${\cal R}(0) \simeq 10^{-3}~{\rm yr^{-1} galaxy^{-1}} {\cal R}_{-3} 
\simeq 2\times 10^{-5}~{\rm yr^{-1} Mpc^{-3}} {\cal R}_{-3}$,
for a number density of galaxies $n_g=0.02~{\rm Mpc}^{-3}$ (Allen 1973).
Assuming that the magnetar birth rate follows the star forming rate,
${\cal R}(z) \simeq {\cal R}(0) (1+z)^3$ for $z<2$ (Lilly et al. 1996). 
The time-dependent beaming parameter $f_b(t)$ (which is the
fraction of magnetars whose neutrino beams are directed towards us, so
the sweep-averaged solid angle of the neutrino beam is $\Delta
\Omega(t)=4\pi f_b(t)$) cancels out. The outer integral is over the
Hubble volume. For remote magnetars, the neutrino flux of an
individual source drops as $D^{-2}$ while the total number of
magnetars increases as $D^3$ for $z\ll1$.  Therefore 
most of the diffuse neutrino emission comes from the farthest
magnetars whose birth rate is the highest.

For young magnetars, the time-dependent neutrino luminosity may be
estimated as in \S3. There are some noticeable differences, however. 
For example, due to radiation reaction and possible pair
screening effect, $\eta_p \ll 1$. On the other hand, the pair
screening altitude could be much higher than the altitude where pions
are generated, so that pions could undergo substantial reacceleration
before decaying. As a result $f_c$ could be $\gg 1$. Notice that
these uncertainties only influence the typical energy of the neutrino 
background, $\bar \epsilon_\nu$, but do not influence the number
counts of the neutrino background (\ref{barphinu}), which can be
estimated as follows. 
The time-dependent neutrino luminosity is 
$L_\nu (t) = A_{pc}(t) c n_{\pi^{+}}(t)\bar \epsilon_\nu$, 
where $A_{pc}(t)=\pi \Omega(t)R^3/c$ is the time-dependent polar cap
area, $\Omega(t) =\Omega_0(1+t/t_{c})^{-1/2}$ is the
time-dependent spin frequency of the magnetar since birth, and
$\Omega_0$ and $t_c$ are constants dependent on the initial rotation
period and polar magnetic field of the magnetar; 
$n_{\pi^{+}} (t)=\xi n_{_{\rm GJ}}(t)=10 \xi_1  n_{_{\rm GJ}}(t)$ is
the time-dependent number density of pions; $\xi \sim R/l_{p\gamma}
\sim 10$ 
is the typical pion multiplicity; $n_{_{\rm GJ}}(t)=\Omega(t) B_p/
2\pi ce$; $\bar \epsilon_\nu$ is the typical neutrino energy whose
detailed value does not enter the problem (i.e. canceled out in
eq.[\ref{barphinu}]).
Averaging over the magnetar neutrino-loud lifetime
$\tau_{mag,\nu}$, and properly taking into account the cosmological
evolution, we estimate
\begin{equation}
\bar \phi_{\nu}
\sim 10^{-13}~{\rm cm^{-2}s^{-1}sr^{-1}}
f_{1/2}\xi_1 {\cal R}_{-3}.
\end{equation}
This background is insensitive to the location of the neutrino death
valley ($\propto \ln \tau_{mag,\nu}$), because logrithmically the
entire magnetar life-time essentially contribute to the final value 
equally. The detectability of this background, however, is sensitively
dependent on the typical energy of the neutrinos, which in turn
depends on whether 
the secondary pions undergo substantial reacceleration before decaying
to neutrinos. This is a difficult problem, which is beyond the scope of
the current paper, and may be addressable by performing detailed
numerical simulations such as those by Harding \& Muslimov (2002). 
Nonetheless, we can set lower and upper bounds for the typical 
neutrino energies. If pion reacceleration is unimportant, the
typical neutrino energy is bound from below to $\sim 2$ TeV according
to the IC cooling argument (\S3), in which case the diffuse background 
is completely masked by the atmospheric background and non-detectable. 
If pion reacceleration is efficient, however, the typical neutrino energy
is bound from above by the radiation reaction limit of the pions, and
the typical neutrino energy could reach 1 PeV or even higher. Such a
neutrino background would become observationally interesting for ICECUBE 
if $\bar \epsilon_\nu \geq 100$ TeV (D. F. Cowen, 2003, private
communication), and at such energies, the diffuse emission from other
neutrino sources becomes weaker than this component (Protheroe 1999). 

\section{Discussion}

We have argued that young magnetars with oppositely oriented magnetic
and spin moments (${\bf \Omega} \cdot {\bf B_p} <0$) can emit high 
energy neutrinos from their polar caps. We point out four neutrino 
emission candidates (Table 1) among the known magnetars. The chances 
of detecting neutrinos from these objects with future km$^3$ neutrino 
telescopes (such as ICECUBE) is small, because their emission is faint
and beamed. Nonetheless, it is finite, and we suggest that these are 
possible discrete neutrino sources for the telescopes to monitor. 
Furthermore, post-burst magnetars (SGRs) should have higher X-ray 
luminosities and hence, should contribute higher neutrino fluxes. This 
enhances the chance of detecting neutrinos from known magnetars.
The level of the diffuse neutrino background contributed by young
magnetars in the whole universe is weak and marginally detectable,
depending on the poorly known typical neutrino energy subject to
further detailed modeling. 

We have concentrated on the possible photomeson interactions near the
magnetar polar cap. However, similar neutrino production processes may 
also take place in the magnetar wind nebula, which could contribute 
additional neutrino emission components besides the one discussed here.

\acknowledgements
This work was supported by NASA NAG5-9192, NSF AST 0098416, 
National Natural Science Foundation of China (grant 19825109)
and National 973 Project of China (NKBRSF G19990754). 
BZ acknowledges
helpful conversations or email communications with C. Thompson,
J. Arons, D. F. Cowen, K. Hurley, B. Gaensler, and T. Montaruli.

\begin{deluxetable}{ccccccc}
\tablecaption{Predicted on-beam neutrino-induced upward muon event
rates for the four potential neutrino-emitting magnetars assuming
they are above photomeson threshold. 
}
\tablewidth{0pt}
\tablehead{
\colhead{Name} & \colhead{$P({\rm s})$}  & \colhead{$\dot
P(10^{-11}{\rm s/s})$} & 
\colhead{ref.} &
\colhead{$B_p(10^{15}{\rm G})$} &
\colhead{$D({\rm kpc})$}
& \colhead{$\frac{dN}{dAdt}({\rm km^{-2}~yr^{-1}})$}}
\startdata
 SGR 1900+14 & 5.16 & 10.9 & [1] &
1.51 & (3.0-9.0) & (1.5-13) (0.1/$\Delta\Omega_\nu$) \\
 SGR 0526$-$66 & 8.04 & 6.6 & [2] &
1.47 & $\sim$ 50 & $\sim$ 0.003 (0.1/$\Delta\Omega_\nu$)\\
1E 1048$-$5937 & 6.45 & 2.2 & [3] &
0.761 & (2.5-2.8) & (0.5-0.7) (0.1/$\Delta\Omega_\nu$) \\ 
 SGR 1806$-$20 & 7.48 & 2.8 & [1] &
0.924 & (13.0-16.0) & (0.01-0.02) (0.1/$\Delta\Omega_\nu$) 
\enddata


\tablecomments{References for the spin parameters.
[1] Hurley 2000 and references therein; [2] Kulkarni et al. 2003; [3]
Mereghetti 2001 and references therein.}

\end{deluxetable}

\begin{figure}
\plotone{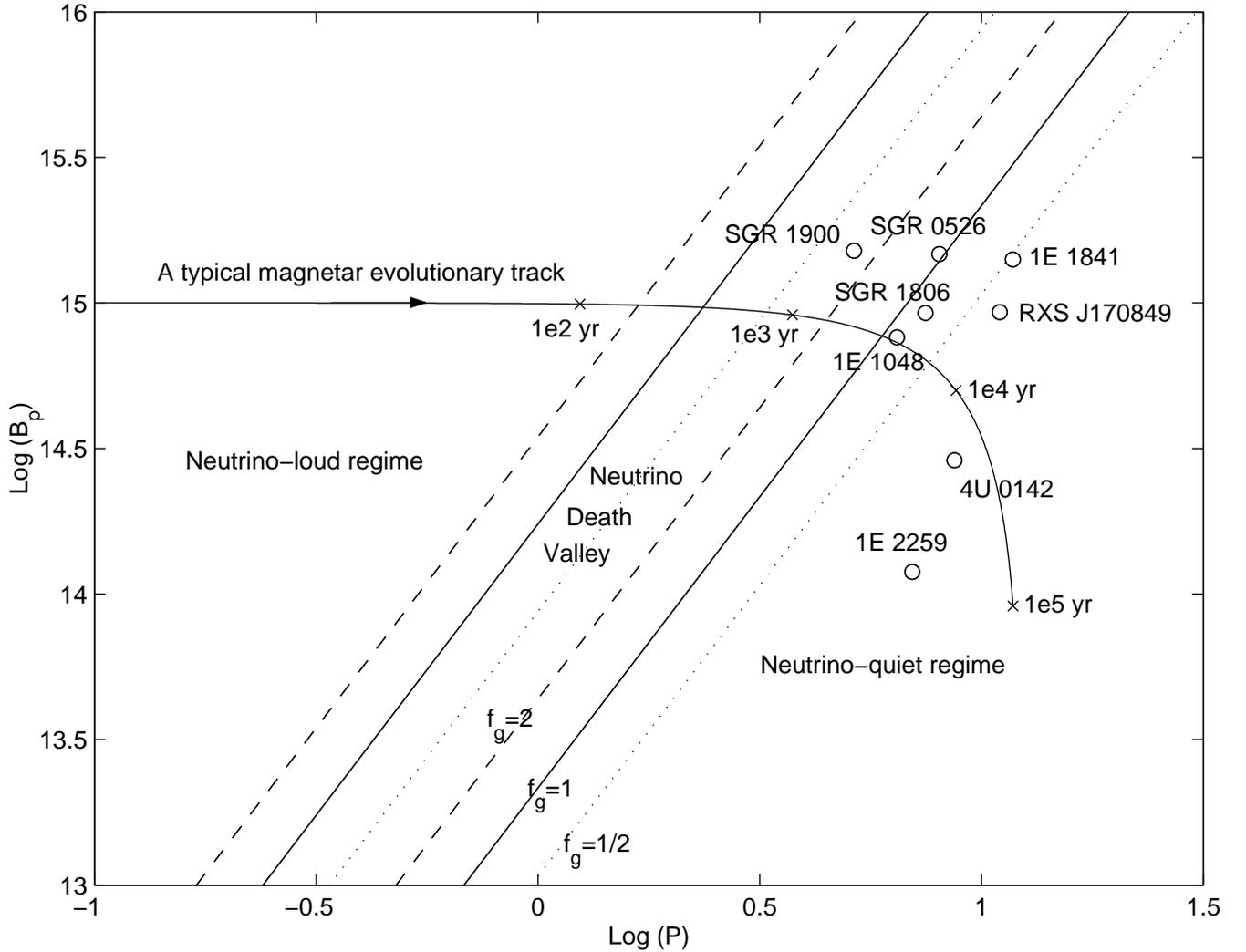} 
\caption{ $P-B_p$ diagram of the known magnetars
with $P$ and $\dot P$ data available (data taken from
http://www.atnf.csiro.au/people/pulsar/catalogue, maintained by
R. N. Manchester; and Kulkarni et al. 2003 for SGR 0526-66) showing
also the neutrino death valley between the two diagonal lines (solid
lines for $f_g=1$; dotted lines for $f_g=1/2$; and dashed lines for
$f_g=2$, where $f_g$ is the angular correction factor for the
threshold condition), and a typical magnetar evolutionary track with
$\tau_{mag} \sim 10^4$ yr (with typical ages marked along the track
with crosses).
 \label{fig1}}
 \end{figure}


\begin{references}

\reference{} Allen, C. W., 1973, Astrophysical Quantities, The
 Athlone Press, London \& Atlantic Highlands

\reference{} Arons, J. 2003, ApJ, in press (astro-ph/0208444)

\reference{} Arons, J. \& Scharlemann, E. T. 1979, ApJ, 231, 854

\reference{} Atoyan, A., \& Dermer, C. D. 2001, Phys. Rev. Lett., 
 87, 221102 

\reference{} Baring, M. G. \& Harding, A. K. 2001, ApJ, 547, 929

\reference{} Beall, J. H., \& Bednarek, W. 2002, ApJ, 569, 343

\reference{} Bednarek, W. 2001, A\&A, 378, L49

\reference{} Blasi, P., Epstein, R. I., \& Olinto, A. V. 2000,
 ApJ, 533, L123

\reference{} Cheng, A. F. \& Ruderman, M. A. 1977, ApJ, 214, 598

\reference{} Colpi, M., Geppert, U. \& Page, D. 2000, ApJ, 529,
 L29


\reference{} Dai, Z. G. \& Lu, T. 2001, ApJ, 551, 249

\reference{} Duncan, R. C. \& Thompson, C. 1992, ApJ, 392, L9

\reference{} Eichler, D. 1978, Nature, 275, 725

\reference{} Goldreich, P. \& Julian, W. H. 1969, ApJ, 157, 869

\reference{} Goldreich, P. \& Reisenegger, A. 1992, ApJ, 395, 250

\reference{} Guetta, D., Spada, M., \& Waxman, E. 2001, ApJ, 559, 101

\reference{} Halzen, F. \& Hooper, D. 2002, Rep. Prog. Phys., 65,
1025

\reference{} Harding, A. K., Baring, M. G. \& Gonthier, P. L.
1997, ApJ, 476, 246

\reference{} Harding, A. K. \& Muslimov, A. G. 1998, ApJ, 508, 328

\reference{} -----. 2001, 556, 987

\reference{} -----. 2002, 568, 862 

\reference{} Heitler, W. 1954, Quantum Theory of Radiation, (Oxford:
Oxford University Press, 1954)

\reference{} Helfand, D. J. 1979, Nature, 278, 720

\reference{} Heyl, J. S. \& Kulkarni, S. R. 1998, ApJ, 506, 61

\reference{} Hurley, K. 2000, in Proc. Fifth Compton Symp.,
Portsmouth, NH, USA, Sep. 1999, AIP Conf. Proc. 510
(eds. M. L. McConnel \& J. M. Ryan), 515 

\reference{} Ibrahim, A. I., Safi-Harb, S., Swank, J. H., Parke, W. et 
al. 2002, ApJ, 574, L51

\reference{} Ibrahim, A. I., Swank, J. H., \& Parke, W. 2003, ApJ,
584, L17 

\reference{} Kaplan, D. L. et al. 2001, ApJ, 556, 399



\reference{} Kulkarni, S. R. et al. 2003, ApJ, 585, 948

\reference{} Levinson, A. \& Waxman, E. 2001, Phys. Rev. Lett., 87,
171101 

\reference{} Lilly, S. J. et al. 1996, ApJ, 460, L1

\reference{} Marsden, D., Lingenfelter, R. R., Rothschild, R. E.
\& Higdon, J. C. 2001, ApJ, 550, 397

\reference{} Mereghetti, S. 2001, in The neutron star - black hole
connection, Proc. of the NATO Advanced Study Institute, Elounda,
Crete, Greece, 7-18 June 1999 (eds. C. Kouveliotou, J. Ventura, \& E.
Van den Heuvel), 351

\reference{} M\'esz\'aros, P., \& Waxman, E. 2001, Phys. Rev. Lett.,
87, 171102

\reference{} Pacz\'ynski, B. 1992, AcA, 42, 145


\reference{} Protheroe, R. J. 1999, Nucl. Phys. B Proc. Supp., 77, 465 

\reference{} Ruderman, M. A. \& Sutherland, P. G. 1975, ApJ, 196,
51

\reference{} Sanwal, D., Pavlov, G. G., Zavlin, V. E. \& Teter, M.
A. 2002, ApJ, 574, L61

\reference{} Stecker, F. W., Done, C., Salamon, M. H. \& Sommers, P.,
1991, Phys. Rev. Lett., 66, 2697

\reference{} Thompson, C. 2001, in The neutron star - black hole
connection, Proc. of the NATO Advanced Study Institute, Elounda,
Crete, Greece, 7-18 June 1999 (eds. C. Kouveliotou, J. Ventura, \& E.
Van den Heuvel), 369

\reference{} Thompson, C. \& Duncan, R. C. 1995, MNRAS, 275, 255

\reference{} -----. 1996, ApJ, 473, 322

\reference{} Thompson, C., Lyutikov, M., \& Kulkarni, S. R. 2002,
ApJ, 574, 332

\reference{} Usov, V. V. 1992, Nature, 357, 472

\reference{} Wang, F. Y.-H., Ruderman, M., Halpern, J. P. \& Zhu,
T. 1998, ApJ, 498, 373


\reference{} Waxman, E. \& Bahcall, J. N. 1997, Phys. Rev. Lett., 78,
 2292 

\reference{} -----. 2000, ApJ, 541, 707

\reference{} Woods, P. M. 2003, to appear in "High Energy Studies of
Supernova Remnants and Neutron Stars" (COSPAR 2002), (astro-ph/0304372)

\reference{} Zavlin, V. E., Pavlov, G. G. \& Tr\'umper, J. 1998,
A\&A, 331, 821

\reference{} Zhang, B., \& Harding, A. K. 2000, ApJ, 532, 1150

\reference{} Zhang, B., Harding, A. K. \& Muslimov, A. G. 2000,
ApJ, 531, L135


\end{references}
\end{document}